\documentclass[]{article}
\usepackage{multicol}
\usepackage{apj}

\input epsf.sty

\newcommand{\mincir}{\raise -2.truept\hbox{\rlap{\hbox{$\sim$}}\raise5.truept
\hbox{$<$}\ }}
\newcommand{\magcir}{\raise -2.truept\hbox{\rlap{\hbox{$\sim$}}\raise5.truept
\hbox{$>$}\ }}
\newcommand{\siml}{\raise -2.truept\hbox{\rlap{\hbox{$\sim$}}\raise5.truept
\hbox{$<$}\ }}
\newcommand{\simg}{\raise -2.truept\hbox{\rlap{\hbox{$\sim$}}\raise5.truept
\hbox{$>$}\ }}

\newcommand{\be}{\begin{equation}}
\newcommand{\ee}{\end{equation}}
\newcommand{\ba}{\begin{eqnarray}}
\newcommand{\ea}{\end{eqnarray}}
\newcommand{\brr}{\begin{array}}
\newcommand{\err}{\end{array}}
\newcommand{\bc}{\begin{center}}
\newcommand{\ec}{\end{center}}
\newcommand{\lb}{{\left<\right.}}
\newcommand{\rb}{{\left.\right>}}
\newcommand{\hm}{\,h^{-1}{\rm Mpc}}
\newcommand{\vel}{\,{\rm km\,s^{-1}}}
\newcommand{\fl}{\,{\rm erg\,s^{-1}cm^{-2}}}
\newcommand{\lum}{\,{\rm erg\,s^{-1}}}

\begin{document}

%%%%%%%%%%%% per formato preprint
\vspace{15mm}
\begin{center}
\uppercase{Cosmological Constraints from the ROSAT Deep Cluster Survey}\\
\vspace*{1.5ex}
{\sc Stefano Borgani$^1$, Piero Rosati$^{2}$, Paolo Tozzi$^{3,4}$ \&
Colin Norman$^4$}\\
\vspace*{1.ex}
{\small
$^1$ INFN, Sezione di Perugia, c/o Dipartimento di Fisica
  dell'Universit\`a, via A. Pascoli, I-06123 Perugia (Italy)\\
E-mail: stefano.borgani@perugia.infn.it\\
$^2$ ESO -- European Southern Observatory, D-85748 Garching
  bei M\"unchen (Germany)\\
E-mail: prosati@eso.org \\
$^3$ Dipartimento di Fisica, II Universit\`a di Roma, via
  Ricerca Scientifica 1, I-00133 Roma (Italy)\\
$^4$ Department of Physics and Astronomy, The Johns Hopkins
  University, Baltimore, MD 21218 (U.S.A.) \\
E-mail: tozzi, norman@stsci.edu\\
}
\end{center}

\vspace*{-6pt}

\begin{abstract}
  The ROSAT Deep Cluster Survey (RDCS) has provided a new large deep
  sample of $X$--ray selected galaxy clusters. Observables such as the
  flux number counts $n(S)$, the redshift distribution $n(z)$ and the
  $X$-ray luminosity function (XLF) over a large redshift baseline
  ($z\mincir 0.8$) are used here in order to constrain cosmological
  models. Our analysis is based on the Press--Schechter approach, whose
  reliability is tested against N--body simulations.  Following a
  phenomenological approach, no assumption is made {\em a priori} on
  the relation between cluster masses and observed $X$--ray
  luminosities.  As a first step, we use the local XLF from RDCS,
  along with the high--luminosity extension provided by the XLF from
  the Brightest Cluster Survey, in order to constrain the amplitude of
  the power spectrum, $\sigma_8$, and the shape of the local
  luminosity--temperature, $L_{bol}$--$T$, relation.  We obtain
  $\sigma_8=(0.58\pm 0.06)\times \Omega_0^{-0.47+0.16\Omega_0}$ for
  flat models ($\Omega_\Lambda=1-\Omega_0$) and $\sigma_8=(0.58\pm
  0.06)\times \Omega_0^{-0.53+0.27\Omega_0}$ for open models
  ($\Omega_\Lambda=0$) at 90\% confidence level, almost independent of
  the $L_{bol}$--$T$ shape.  The density parameter $\Omega_0$ and
  the evolution of the $L_{bol}$--$T$ relation are constrained by
  the RDCS XLF at $z>0$ and the EMSS XLF at $\bar z=0.33$, and by the
  RDCS $n(S)$ and $n(z)$ distributions. By modelling the evolution for
  the amplitude of the $L_{bol}$--$T$ relation as $(1+z)^A$, an
  $\Omega_0=1$ model can be accommodated for the evolution of the XLF with
  $1\mincir A\mincir 3$ at 90\% confidence level, while
  $\Omega_0=0.4^{+0.3}_{-0.2}$ and $\Omega_0\mincir 0.6$ are implied
  by a non--evolving $L_{bol}$--$T$ ($A=0$) for open and flat models,
  respectively.

\vspace*{6pt}
\noindent
{\em Subject headings: } 
Cosmology: theory - dark matter - galaxies: clusters:
general - X-rays: galaxies

\end{abstract}

%%%%%%%%%%%% per formato preprint
\begin{multicols}{2}
%%%%%%%%%%%%

\section{Introduction}

Galaxy clusters are crucial probes for models describing the formation
and evolution of cosmic structures.  In standard scenarios, clusters
form at high peaks of the primordial density field (e.g., Kaiser 1984;
Bardeen et al. 1986). Therefore, both the statistics of their
large--scale distribution and their abundance are highly sensitive to
the nature of the underlying dark matter density field.  Furthermore,
their typical scale, $\sim\! 10 \hm$ ($h$ is the Hubble constant in
units of $100\vel Mpc^{-1}$), relates to fluctuation modes which have
still to enter, or are just approaching, the non--linear stage of
gravitational evolution. Then, although their internal gravitational
and gas dynamics are rather complex, a statistical description of
global cluster properties can be provided by resorting to linear
theory or perturbative approaches.  By following the redshift
evolution of clusters, we have a valuable method to trace the global
dynamics of the Universe and, therefore, to determine its geometry.

In this context, the cluster abundance at a given mass has long been
recognized as a stringent test for cosmological models. Typical rich
clusters have masses of about $5\times 10^{14}h^{-1}M_\odot$, which is
quite similar to the average mass within a sphere of $\sim 8\hm$
radius in the unperturbed universe.  Therefore, the local abundance of
clusters is expected to place a constraint on $\sigma_8$, the
r.m.s. fluctuation on the $8\hm$ scale.  Analytical arguments based on
the approach devised by Press \& Schechter (1974) show that the cluster
abundance is highly sensitive to $\sigma_8$ for a given value of the
density parameter $\Omega_0$ (e.g., Frenk et al.  1990; Bahcall \& Cen
1992; Lilje 1992; White, Efstathiou \& Frenk 1993; Viana \& Liddle
1996).  Once a model is tuned so as to predict the correct abundance
of local ($z\mincir 0.1$) clusters, its evolution will mainly depend
on $\Omega_0$ (e.g., Oukbir \& Blanchard 1992). Hence, the possibility
of tracing the evolution of the cluster abundance with redshift should
provide both the value of the density parameter and the 
amplitude of the fluctuations at the cluster scale.

It is worth stressing, however, that by `abundance' one means the
comoving volume density of galaxy clusters {\em per unit mass}, which
cannot be directly observed. Therefore, these arguments always assume
a relation of some observable cluster property to the total mass.
In this respect, X-ray selected samples have been proved to
be extremely useful, essentially due to the simplicity of their
selection functions.

The availability of new observational data for high--$z$ clusters
has recently stimulated a flurry of activity in this direction.  For
instance, Carlberg et al. (1997) and Fan, Bahcall \& Cen (1997) have
used results on the internal velocity dispersion $\sigma_v$ of X-ray
selected clusters in the CNOC survey (Carlberg et al. 1996) and
concluded that low--density models with $\sigma_8\simeq 0.7$--0.9 and
$\Omega_0\simeq 0.3$--0.5 are preferred (see, however, Gross et al. 1998).

Observations of cluster X-ray temperatures offer an independent and
powerful means to characterize the evolution of the cluster abundance
since cluster temperatures are directly connected to cluster masses.  Eke,
Cole \& Frenk (1996) have pointed out that a measurement of the
cluster temperature distribution at $z\simeq 0.5$ would discriminate
at a high confidence level between a critical density model and a
low--density model with $\Omega_0\simeq 0.3$. First attempts in this
direction have been pursued by Henry (1998) and Eke et al. (1998) who
used available data on the $X$--ray temperature function for $z\mincir
0.4$ and found $\sigma_8\simeq 0.5$--0.8 and $\Omega_0\simeq
0.3$--0.7. 
Based on a similar data set, a different conclusion has been however 
reached by Viana \& 
Liddle (1998), who argued that 
a critical--density Universe is still viable as far as the evolution of the 
cluster temperature function is concerned.
More robust results will require substantially larger compilations of
cluster temperatures, which will be a time consuming process even with
the next generation of X-ray satellites.

An alternative way to trace the evolution of the cluster abundance is
to rely on the luminosity distribution of X-ray flux--limited cluster
samples (e.g., Henry et al. 1992; Oukbir \& Blanchard 1992; Bartlett
\& Silk 1993; Colafrancesco \& Vittorio 1994; Reichart et al. 1998).
The advantage of this approach lies in the simplicity of the
measurement and the availability of large samples, with well
understood selection biases. A good understanding of hydrodynamical
processes in the intra--cluster medium (ICM), as well as feedback
mechanisms from stellar energy release, are however needed in order to
relate the $X$--ray luminosity to the cluster mass.

Following the original samples from the Einstein Medium Sensitivity
Survey (EMSS; Gioia et al. 1990), the ROSAT-PSPC has recently given a
strong impulse in this area providing both large solid angle, high
flux limit samples (e.g. the Bright Cluster Sample (BCS) by Ebeling et
al. 1997), and more distant cluster samples from deeper, small solid
angle surveys (the ROSAT Deep Cluster Survey (RDCS) by Rosati et
al. 1995, 1998; WARPS by Scharf et al. 1997; SHARC-S by Burke et
al. 1997 and Collins et al. 1997; the sample based on ROSAT PSPC by
Vikhlinin et al. 1998).

In this paper, we use, for the first time, the complete amount of
information contained in the RDCS, namely the evolving $X$--ray
luminosity function (XLF), the deepest flux number counts and the
redshift distribution, along with a knowledge of the RDCS selection
function.  In particular, in order to normalize the models at $z\sim
0$ we will use the local XLF from the RDCS and the high--luminosity
extension from BCS. Then, we will use the XLF estimates from RDCS at
the median redshifts $\bar z \simeq 0.3$ and $\bar z\simeq 0.6$, the
flux--number counts $n(S)$ and the redshift distribution $n(z)$ from
the RDCS to constrain the evolution.  We note that the number counts
and redshift distribution represent the projection of the
$z$--dependent XLF along the redshift and flux, respectively.  The
number counts alone from the RDCS have been already used by different
authors to place constraints on cosmological models (see Kitayama \&
Suto 1997; Mathiesen \& Evrard 1998 for a parametric approach; see
Cavaliere, Menci \& Tozzi 1998 for an approach based on a physical
models for the ICM).

Instead of relying on specific physical models to describe the ICM and
its evolution, we prefer here to adopt a phenomenological approach.
Using a parameteric expression for the relation between cluster
masses and $X$--ray luminosities, we fit the corresponding parameters,
along with those describing the cosmological models, to match the RDCS
data.

The two principal issues addressed in this paper are to: (1) derive
robust constraints on the amplitude of fluctuations on the cluster
mass scale using the local XLF and; (2) derive robust constraints on
$\Omega_0$ and the $L_{bol}$--$T$ relation from a flux--limited sample
of clusters with $z\mincir 1$, like RDCS.

The structure of the paper is as follows.  In Section 2 we give a
brief description of the data, while in Section 3 we present the
method of analysis. We review the Press--Schechter approach for the
mass function and compare its predictions with a set of P3M N--body
simulations in order to assess its reliability. Then, we describe in
detail our procedure to convert cluster masses into luminosities.
This will be accomplished in three steps: {\em (i)} converting virial
masses into temperatures; {\em (ii)} converting temperatures into
bolometric luminosities; {\em (iii)} correcting luminosities from the
bolometric to the soft {\sl ROSAT} [0.5-2.0] keV band.  In Section 4
we present the results of this analysis from the local XLF to
constrain the shape and the amplitude $\sigma_8$ of {\sl
COBE}--normalized power--spectra as well as the local
luminosity--temperature ($L_{bol}$--$T$) relation.  In Section 5 we
use the evolving XLF, number counts and redshift distribution to
constrain the density parameter $\Omega_0$ and the evolution of
$L_{bol}$--$T$. A brief discussion of the results and the main
conclusions are given in Section 6.

\section{The RDCS sample}
The RDCS compiled a large, X-ray flux limited sample of
galaxy clusters, selected on the basis of X-ray properties alone, via
a serendipitous search in ROSAT-PSPC deep pointed observations (Rosati
et al. 1995).  The depth and solid angle of the survey were chosen 
to probe an adequate range of X-ray luminosities over a large
redshift baseline.  Over 160 cluster candidates were selected down to
the flux limit of $S=1\times 10^{-14}\fl$, over an area of 50 deg$^2$,
by utilizing a wavelet-based detection technique. The latter is
particularly efficient in discriminating between point-like and
extended, low surface brightness sources.  The completeness and degree
of contamination of the catalogue are well understood above $S=2\times
10^{-14}\fl$, but still uncertain at lower fluxes where the optical
identification program is not yet complete. The sky coverage of the
survey, a crucial ingredient in the interpretation of deep cluster
surveys, has also been extensively studied (Rosati et al. 1998; RDNG
hereafter).  Cluster redshifts have been secured for more than 100
clusters/groups to date using NOAO and ESO telescopes. In this paper,
we use the flux limited subsample of spectroscopically identified
clusters used by RDNG to derive the cluster XLF.  This comprises 70
clusters with fluxes $S\ge 4\times 10^{-14}\fl$ and $0.05\siml z\siml
0.85$. The whole RDCS sample is used to compare model predictions with
the observed cluster number counts.  An additional determination is
included in the present analysis at the faintest bin ($S<2\times
10^{-14}\fl$) which extends the Log N--Log S presented by RDNG.  We
note that 15\% of the cluster candidates, mostly at the faint end,
still remain to be optically identified.

\section{The Theoretical framework}

\subsection{The recipe for the cluster mass function}

\subsubsection{The power spectrum model}
We write the linear power spectrum of density fluctuations as
$P(k)\propto k^{n_{pr}}T^2(k)$. For the transfer function $T(k)$,
we assume the Gamma--model
\ba
& &T(q)\, = \, {{\rm ln}(1+2.34 q)\over 2.34 q}\times \nonumber \\
& &\left[1+3.89q+(16.1q)^2+(5.46q)^3+(6.71q)^4\right]^{-1/4}
\label{eq:tk}
\ea 
where $q=k/h\Gamma$, being $\Gamma$ the shape parameter. For the
class of CDM models, it is $\Gamma\simeq \Omega_0h$ (Bardeen et al. 1986),
while in general $\Gamma$ can be viewed as a free parameter, to be
fitted to observational constraints. For instance, for the
$\Omega_0=1$, case $\Gamma\simeq 0.2$ can be obtained either in the
framework of Cold+Hot DM models with a suitable choice of the massive
neutrino fraction (e.g., Primack 1997), or in the framework of
$\tau$CDM models, where the CDM shape of $P(k)$ is modified by the
decay of massive neutrinos (White, Gelmini \& Silk 1995). As for the
primordial spectral index, $n_{pr}$, we will mainly concentrate in the
following on the Harrison--Zeldovich case $n_{pr}=1$, although we will
also comment about the effect of varying $n_{pr}$ around this value.
As usual, the amplitude of $P(k)$ will be expressed in terms of
$\sigma_8$ and is fixed by the four--year {\sl COBE}
normalization recipe, as provided by Bunn \& White (1997) and Hu \&
White (1997) for the $\Omega_\Lambda=1-\Omega_0$ and the 
$\Omega_\Lambda=0$ cases, respectively. This normalization determines
a one-to-one correspondence between $\sigma_8$ and the shape parameter
$\Gamma$ for a fixed value of $n_{pr}$.

\subsubsection{The Press--Schechter approach}
According to the PS formalism, clusters at a given redshift, $z$, are
identified with those halos that are just virializing. The
comoving number density of such structures in the mass range
$[M,M+dM]$ reads
\be
{dn\over dM}\,=\,\sqrt{2\over \pi}\, {\bar \rho \over M^2}\,
{\delta_c(z)\over \sigma_M}\, \left|{d\log \sigma_M\over d\log
M}\right|\, \exp\left(-{\delta_c(z)^2\over 2\sigma_M^2}\right)\,.
\label{eq:ps}
\ee
Here $\bar \rho$ is the present day average matter density and 
$\delta_c(z)$ is 
the linear--theory overdensity extrapolated at the present time 
for a uniform spherical fluctuation collapsing at redshift $z$. This
quantity conveys
information about the dynamics of fluctuation evolution in a generic
Friedmann background. It is convenient to express it as
$\delta_c(z)=\delta_0(z)\,[D(0)/D(z)]$, where 
\be
D(z)\,=\,{5\over 2}\,\Omega_0 E(z)\,\int_z^\infty {1+z'\over
E(z')^3}\, dz'
\label{eq:grw}
\ee 
is the linear fluctuation growth factor. In the above expression,
$E(z)=[(1+z)^3 \Omega_0+(1+z)^2\Omega_R+\Omega_\Lambda]^{1/2}$, where
$\Omega_\Lambda =\Lambda/3H^2$ and
$\Omega_R=1-\Omega_0-\Omega_\Lambda$ (see, e.g., Peebles 1993). 

The quantity $\delta_0(z)$ has a weak dependence on $\Omega_0$ for
both $\Omega_\Lambda=0$ and $\Omega_R=0$. In the following we will
adopt for $\delta_0(z)$ the expression provided by Kitayama \& Suto
(1996). For a critical--density Universe it is $\delta_c(z)=1.686(1+z)$.

The rms density fluctuation at the mass scale $M$, $\sigma_M$,  is
connected to the fluctuation power spectrum according to
\be
\sigma^2_M\,=\,{1\over 2\pi^2}\,\int_0^\infty dk\,k^2\,P(k)\,W^2(kR)\,.
\label{eq:sigm}
\ee
Here $W(x)$ is the Fourier representation of the window function,
which describes the shape of the volume from which the collapsing
object is accreting matter. The comoving fluctuation size $R$ is
connected to the mass scale $M$ as $R=(3M/4\pi \bar\rho)^{1/3}$ for the
top--hat window, $W(x)=3(\sin x- x\cos x)/x^3$, that we adopt in the
following.

\subsubsection{Comparison with N--body simulations}
The reliability of the PS formula has been tested against N--body
simulations by several authors. As a general result, it turns out that
eq.(\ref{eq:ps}) provides an overall satisfactory description of the
N--body mass function around the non--linear mass scale $M_*$ (i.e.,
the mass at which $\delta_c(z)/\sigma_{M_*}=1$), which controls the
position of the exponential cut--off (e.g., Lacey \& Cole 1993; Eke et
al.  1996).

In order to make our own comparison, we ran a set of N--body
simulations for three different models.  Simulations are run for: {\em
(a)} an $\Omega_0=1$ model with $n_{pr}=0.8$ and $\Gamma=0.35$; {\em
(b)} a flat low--density model with $\Omega_0=0.4$ and $\Gamma=0.22$;
{\em (c)} an open model with $\Omega_0=0.6$ and $\Gamma=0.25$. Model
parameters are described in Table 1. They have been chosen
in such a way that the corresponding power spectra are consistent with
the APM galaxy $P(k)$ shape (e.g., Baugh \& Efstathiou 1993), the
local cluster number density (e.g., Eke et al. 1996; Girardi et
al. 1998) and the CMB anisotropies from the 4--year {\sl COBE} data
(e.g., Gorsk\`{\i} et al. 1996).

Each simulation is run within a box of ${\cal L}=250\hm$ a side using
the adaptive particle-particle--particle-mesh (AP3M) code, kindly
provided by Couchman (1991). The code follows the trajectories of
$128^3$ particles, with a comoving equivalent Plummer softening
parameter of $\simeq 100\,h^{-1}$ kpc.  Therefore, both the mass
resolution ($m_{part}\simeq 2.06\times 10^{12}\Omega_0\,h^{-1}M_\odot$
for the mass of each particle) and the dynamical resolution are
adequate to describe halo masses down to poor
clusters ($\mincir 10^{14}h^{-1}M_\odot$).  The initial redshift $z_i$
at which simulations are started has been fixed so that $\sigma=0.2$
for the r.m.s. fluctuation amplitude on the grid at that time. The
integration variable is chosen to be $p=a^{3/2}$, where $a=(1+z)^{-1}$
is the expansion factor. The number of time--steps has been determined
by fixing $\Delta p=0.14$. This ensured an energy conservation of
about 3\% at the final step of each simulation.  For each model, we
run five different realizations in order to have enough statistics to
reliably estimate the mass function in the high--mass tail.

%%TAB1
\vspace{6mm}
\hspace{-4mm}
\begin{minipage}{9cm}
\renewcommand{\arraystretch}{1.2}
\renewcommand{\tabcolsep}{1.2mm}
\begin{center}
\vspace{-3mm}
TABLE 1\\
\vspace{2mm}
{\sc Parameters of the simulation models\\}
\footnotesize
\vspace{2mm}

\begin{tabular}{lcccccc}
\hline \hline
\multicolumn{1}{l}{Model}
&\multicolumn{1}{c}{$\Omega_0$}
&\multicolumn{1}{c}{$n_{pr}$}
&\multicolumn{1}{c}{$\Gamma$}
&\multicolumn{1}{c}{$\sigma_8$}
&\multicolumn{1}{c}{$z_i$}
&\multicolumn{1}{c}{Time--steps}
\\
\hline
EdS           & 1.0 & 0.8  & 0.35 & 0.56 & 10 & 240 \\
Flat          & 0.4 & 1.0  & 0.22 & 0.87 & 16 & 480 \\
Open          & 0.6 & 1.0  & 0.25 & 0.67 & 15 & 440 \\

\hline
\end{tabular}
 
\end{center}
\vspace{3mm}
\end{minipage}
\normalsize

%%FIGURE 1%%%
%\end{multicols}
%\begin{figure}
\includegraphics{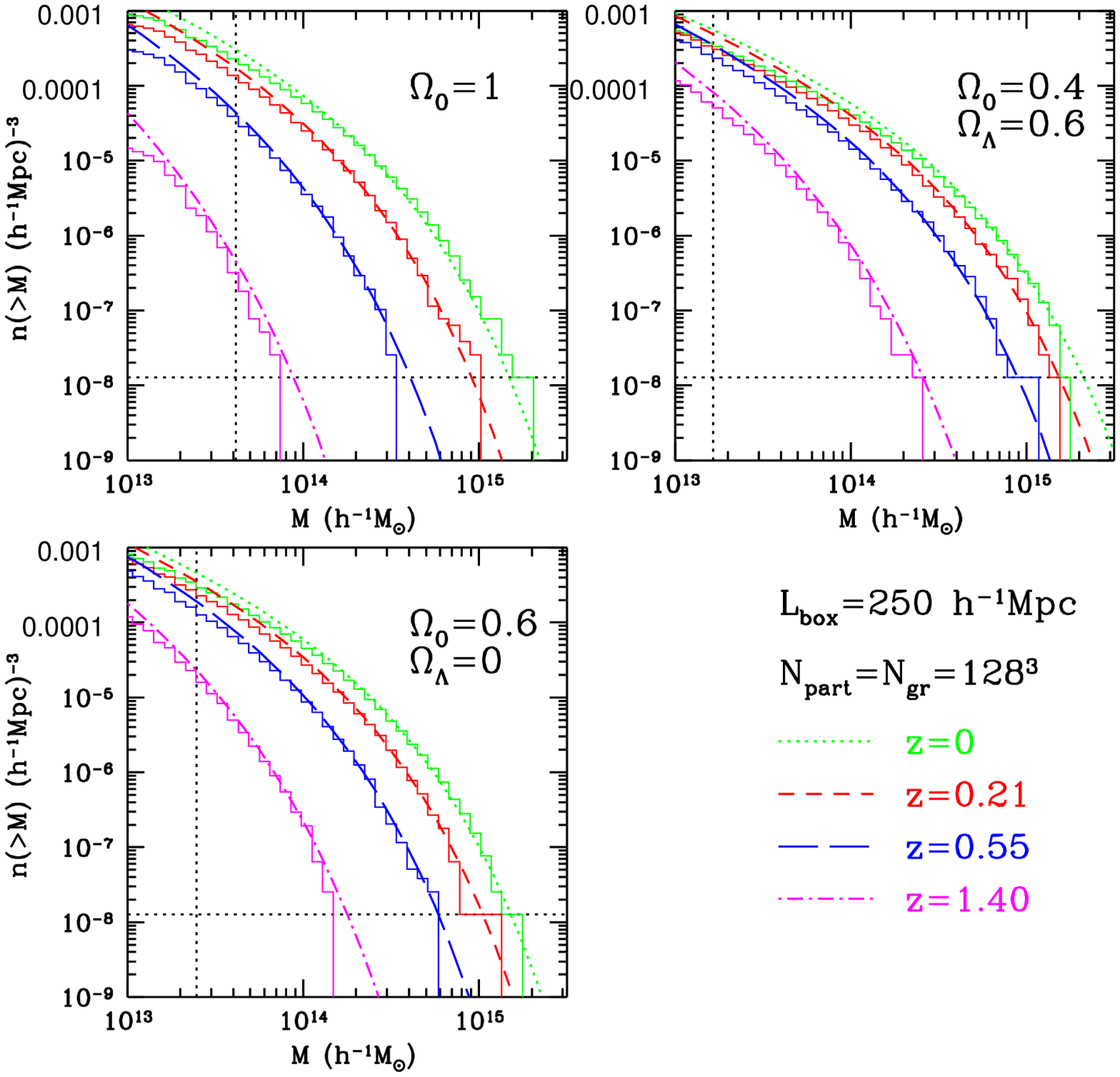}
%\special{psfile=f1.eps angle=0 voffset=-330 hoffset=-10 vscale=45 hscale=45}
$\ \ \ \ \ \ $\\
\vspace{8.2truecm}
$\ \ \ $\\
%\vspace{-0.5truecm}
{\small\parindent=3.5mm {Fig.}~1.--- 
Comparison between the cumulative cluster mass
  function $n(>M)$ produced by N-body simulations (histograms) and the
  Press--Schechter formula (smooth curves) at different
  redshifts. Details about the parameters of the simulated models are
  provided in Table 1. The horizontal dotted line indicates the
  shot-noise level, i.e. the cluster number density associated with one
  single cluster in the volume of the five simulation boxes. The
  vertical dotted lines indicates the minimum fiducial mass for a
  cluster to be reliably identified, i.e.  corresponding to a cluster
  resolved with 20 particles.}
%\label{fi:mfps}
%\end{figure}

%\begin{multicols}{2}
%%%%%%%%%%%%%%

Clusters in each simulation are identified by following a two--step
procedure. Firstly, we apply a friend-of-friends (FOF) algorithm with
linking parameter $b=0.2$ in order to identify cluster candidates. For
each FOF group, we compute the center of mass of the particles that
belong to it. Therefore, starting from each center, draw a sphere
whose radius encompasses an average density equal to the virial density:
$\rho_{vir}(z)= \Delta_{vir}(z)\,\bar \rho(z)$ where $\bar \rho (z)$
is the cosmic average density at redshift $z$. 
It is $\Delta_{vir}=18\pi^2$ for $\Omega_0=1$,
whereas we take the expressions provided by Kitayama \& Suto (1996)
for the $\Omega_0<1$ cases. The center of mass of all the particles
within such a sphere is then computed and used as the new
center for the next iteration. The procedure is stopped, in general
after few iterations, once the coordinates of the center of the sphere
and its mass converge.  When two clusters are found to partially
overlap, the smaller one is discarded from the final list.

In Figure 1 we plot the cumulative mass function $n(>M)$,
defined as the number density of objects with masses larger than $M$.
Simulation results (histograms) are compared to the PS predictions
(dashed curves) at four different redshifts, $z=0,0.21,0.55,1.40$. In
each panel, the horizontal dotted line indicates the shot--noise
level -- the number density corresponding to finding one single
cluster in all the five boxes. The vertical dotted lines indicate the
mass of a halo containing 20 particles, smaller halos being not
properly resolved.

As a general result, we confirm that the PS approach provides a good
approximation to the N--body mass function in the region around
the non--linear mass scale $M_*$, quite independently of the
model. The ability of the PS formula to account for the different
degrees of evolution in the different cases is remarkable.
A similar result about the reliability of the PS formula has been also
recently obtained from the analysis of Hubble--volume simulations
(White 1998, private communication; see also Colberg et al. 1998).
Governato et al. (1998) analysed two high--resolution simulations, for
critical density and low--density CDM models, which were aimed at
testing the PS formula. They found a slightly different redshift
dependence of $\delta_c$, which turns into a $\sim 6\%$ smaller value
at $z\simeq 0.5$ only for the $\Omega_0=1$ case. Even in this case,
such a difference only induces variation in $\sigma_8$ and $\Omega_0$
which are smaller than the statistical uncertainties of our analysis
(see Section 4, below).

At low masses the PS expression provides a systematic overestimate of
the halo number density (see also Bryan \& Norman 1998; Gross et
al. 1998).  However, given the flux--limit of the RDCS sample, such
small masses enter only at the lowest redshifts, whereas we will
mostly use the observed distributions at high--redshifts ($z\magcir
0.3$) to constrain our models (we will further comment in the
following on the mass scales associated to RDCS clusters at different
redshifts). 

\subsection{The mass--temperature relation}
According to the spherical collapse model and under the assumption of virial
equilibrium and isothermal gas distribution, the mass--temperature
relation can be written as
\ba
k_BT & = & {1.38\over \beta}\,\left({M\over
10^{15}h^{-1}M_\odot}\right)^{2/3}\nonumber \\
& \times & \left[\Omega_0\Delta_{vir}(z)\right]^{1/3} \,(1+z)\,{\rm keV}\,,
\label{eq:mt}
\ea
where 76\% of the gas is assumed to be hydrogen (see, e.g., Eke et
al. 1996). The 
$\beta$ parameter is defined as the ratio of the specific kinetic energy of
the collisionless matter to the specific thermal energy of the gas,
\be
\beta\,=\,{\mu m_p\sigma_v^2\over k_BT}\,,
\label{eq:beta}
\ee
being $\mu=0.59$ the mean molecular weight, $m_p$ the proton mass and
$\sigma_v$ the one--dimensional cluster internal velocity dispersion.
Observational data on clusters with reliable determinations of both
X-ray gas temperature and galaxy velocity dispersion indicate $\beta
\simeq 1$ for $T\magcir 3$ keV, (e.g., Lubin \& Bahcall 1993; Girardi
et al. 1996). The calibration of the $\beta$ value
using numerical simulations have been attempted by several authors
(see, e.g., Bryan \& Norman 1998, for a summary of numerical results),
by fitting either the $M$--$T$ relation of eq.(\ref{eq:mt}) or the
$T$--$\sigma_v$ relation of eq.(\ref{eq:beta}). Differences between
the resulting $\beta$ values could be due to a non--perfect
virialization of the cluster or to a departure from the hydrostatic
equilibrium in the ICM.  Here we will refer to results based on the
$M$--$T$ relation, that we will use to convert masses into X-ray
luminosities.

Using a Smoothed Particle Hydrodynamics (SPH) simulations of six
clusters, Evrard (1991) found $\beta=1.23$ by adopting a
mass--weighted temperature definition. Evrard, Metzler \& Navarro
(1996) analyzed an enlarged sample of clusters, also including
low--$\Omega_0$ cases, and found an average value of $\beta=1.05$.
Their results also show some dependence on $\Omega_0$, low--density
models preferring a lower $\beta$. Bryan \& Norman (1998) analyzed a
set of cluster simulations based on the piecewise--parabolic--mesh
(PPM) method. Using a luminosity--weighted definition for $T$, they
found $1.27 \le \beta \le 1.33$. Pen (1998) used Moving--Mesh
Hydrodynamic (MMH) cluster simulations for different cosmological
models. By assuming an emission--weighted temperature, he found $\beta
=1.12 \pm0.04$. Recent simulations of the Santa Barbara Cluster (Frenk
et al. 1998), based on a variety of numerical techniques, converge to
indicate that $\beta \simeq 1.15$. This value, which also falls within
the range of previous results, will be adopted in the following as the
fiducial one. We will also show results for $\beta=1$.

\subsection{The luminosity--temperature relation}
The observational determination of the relation between bolometric
luminosity and temperature, $L_{bol}$--$T$, at low redshift is
becoming more and more accurate as larger and higher quality data sets
are constructed (e.g., Mushotzky 1984; Edge \& Stewart 1991; David et
al. 1993). If we model the $L_{bol}$--$T$ relation as
\be
L_{bol}\,=\,L_6\,\left(T\over 6{\rm keV}\right)^\alpha (1+z)^A\times
10^{44} h^{-2}\lum \,,
\label{eq:lt}
\ee 
then low redshift data for $T\magcir 3\,$keV indicates
$L_6\simeq 3$ as a rather stable result, and $\alpha \simeq 2.7$--3,
depending on the sample and the data analysis technique. As for the
behavior at lower temperatures, Ponman et al. (1996) analysed a set
of {\sl ROSAT} observations for 22 Hickson's compact groups and found
indications for a steepening of the $L_{bol}$--$T$ relation below 1
keV. White, Jones \& Forman (1997) analysed a set of 207 {\sl
  EINSTEIN} clusters and found $\alpha \simeq 3$. Although
the formal fitting uncertainties are generally small, the scatter of
data points around the relation (\ref{eq:lt}) is so large as to raise
the question of whether it represents a good model for the
observational $L_{bol}$--$T$ relation. In the same paper, White et al.
also confirm the result by Fabian et al. (1994) concerning the dependence
of the scatter in the $L_{bol}$--$T$ relation on the cooling--flow
(CF) mass deposition rate, a smaller scatter being found when
excluding CF clusters. Similar results about a tight $L_{bol}$--$T$ relation
have been obtained also by Arnaud \& Evrard (1998), Markevitch (1998) 
and Allen \& Fabian (1998) by either considering only clusters with a low 
CF or by correcting for the CF effect.

As for the evolution of the $L_{bol}$--$T$ relation, Mushotzky \&
Scharf (1997) compared results from a sample of {\sl ASCA}
temperatures at $z>0.14$ with the low--redshift sample by David et al.
(1993). They found that data out to $z\simeq 0.4$ are consistent with
no evolution (i.e., $A\simeq 0$), although within rather large
uncertainties. Henry (1997) determined the luminosity--temperature
relation for a sample of {\sl EMSS} clusters at a median redshift
$z=0.32$ using {\sl ASCA} temperature measurements. The decrement he
found in the amplitude of the $L_{bol}$--$T$ relation with respect to
that by David et al. (1993) implies a marginally positive evolution
with $A=0.36 \pm 0.32$. Sadat, Blanchard \& Oukbir (1998) analysed a
compilation of $z>0$ clusters taken from different authors and also
found a mildly positive evolving $L_{bol}$--$T$ with $0\mincir
A\mincir 1$.

On the theoretical side, the first attempt to describe the
$L_{bol}$--$T$ relation and its evolution has been developed by Kaiser
(1986) for an Einstein--de Sitter (EdS) model. This model, based on
the assumptions of self--similar gas evolution, can be extended to
generic cosmologies and predicts $L_{bol}\propto
T^2\Delta_{vir}^{1/2}(z)$.  The resulting slope of the local
$L_{bol}$--$T$ is definitely shallower than the observed one. On the
other hand, comparisons between the self--similar scaling and results
from hydrodynamical simulations have shown a general good agreement
(e.g., Eke, Navarro \& Frenk 1998, and references therein), especially
when the effects of finite numerical resolution are taken into account
(e.g., Bryan \& Norman 1998). Therefore, the basic discrepancy between
the predicted $L_{bol}$--$T$ relation and the observed one,
$L_{bol}\propto T^\alpha$ with $\alpha \sim 3$, must be considered as
an open problem whose solution calls for additional physics, like the
preheating generated by non gravitational sources (e.g., Cavaliere,
Menci \& Tozzi 1997; see, however, Metzler \& Evrard 1997).

As for the $z$--dependence of the $L_{bol}$--$T$ relation, Kaiser
(1986) predicted an induced evolution in the XLF, for acceptable
scale--free power spectra, which is opposite to the mild negative
evolution found later in the data (e.g., Gioia et al. 1990). This led
Kaiser (1991) and Evrard \& Henry (1991) to assume that an initial
entropy was imprinted in the ICM.  Afterwards, this model has been
generalized by Bower (1997) to include the general redshift evolution
for the minimal entropy level. As a result, a range of evolutionary
patterns can be derived for $L_{bol}$--$T$ and, therefore, for the
XLF.

In our analysis we prefer to adopt a conservative approach and, {\em
instead of assuming} a unique shape and evolution for the
$L_{bol}$--$T$ relation, we {\em fit} the corresponding parameters
$\alpha$ and $A$ to the observational data. The amplitude of the
$L_{bol}$--$T$ relation is taken to be $L_6=2.9$ which represents the
best--fitting value to the data by White et al. (1997).

In order to convert bolometric luminosities into the soft {\sl ROSAT}
band $[0.5,2.0]\,$keV we need to introduce the appropriate correction.
We perform it by using a Raymond--Smith code and assuming an overall
ICM metallicity of 0.3 times the solar value, as actually observed.
The ratio between the bolometric and the finite--band luminosity,
$L_{bol}/L_{[0.5-2.0]}$, is plotted in Figure 2. In this plot we also
show the differences respect to a pure bremsstrahlung spectrum with a
power--law approximation for the Gaunt factor, $g(E/k_BT)\propto
(E/k_BT)^{-\gamma}$.  Although this simplified model for $\gamma =0.3$
provides a reasonably accurate bolometric correction for rich
clusters, it becomes inadequate for $T\mincir 2\,$keV where the effect
of metal emission lines in the cluster spectra start playing a
non--negligible role, increasing the emissivity above the
Bremsstrahlung prediction (Sarazin 1988).

%%FIGURE 2%%%
%\end{multicols}
%\begin{figure}
\includegraphics{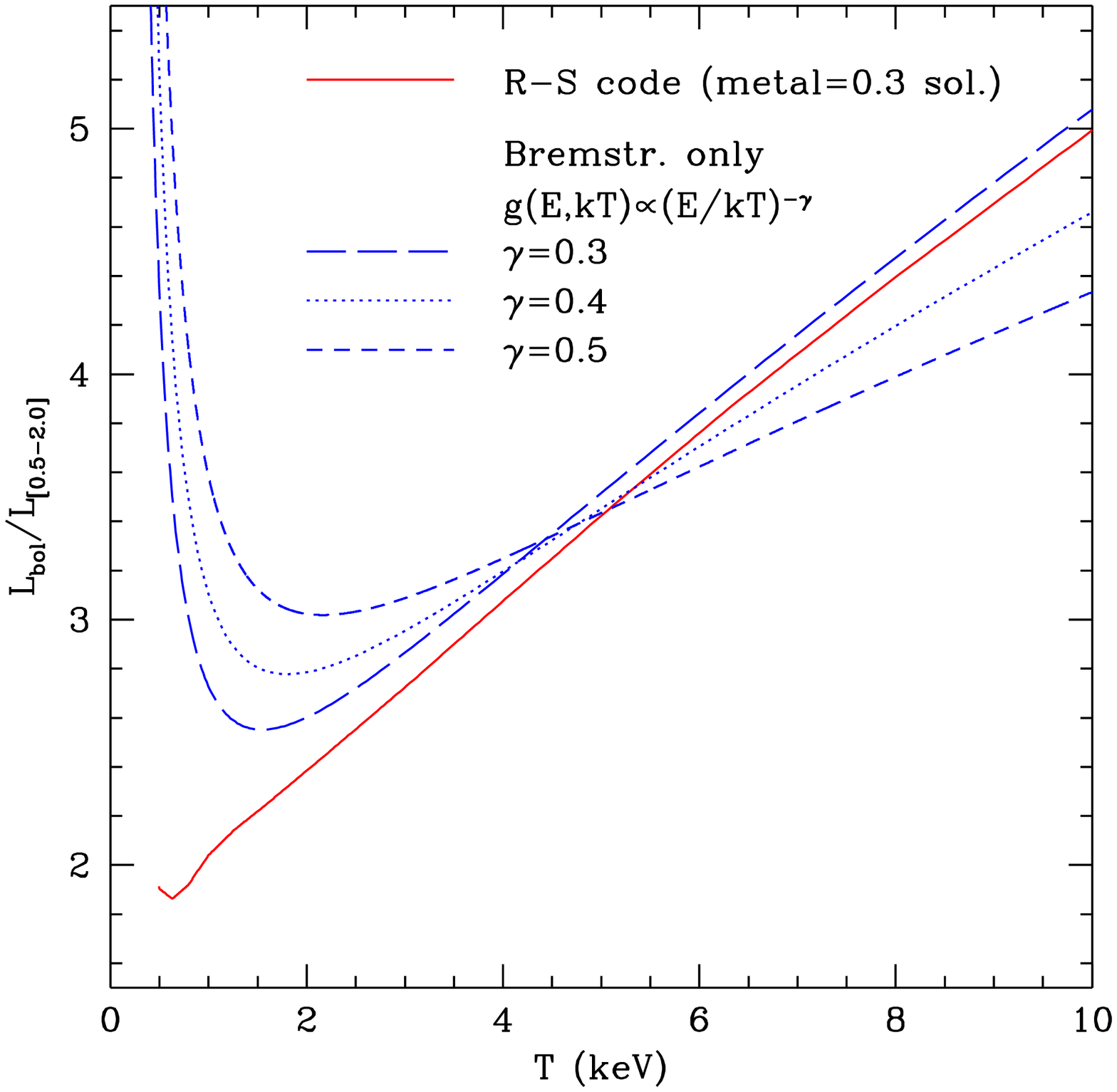}
%\special{psfile=f2.eps angle=0 voffset=-330 hoffset=-10 vscale=45 hscale=45}
$\ \ \ \ \ \ $\\
\vspace{8.2truecm}
$\ \ \ $\\
%\vspace{-0.5truecm}
{\small\parindent=3.5mm {Fig.}~2.--- 
The bolometric correction, as computed from a
  Raymond--Smith code with ICM metallicity of $0.3Z_\odot$
  (solid curve), compared with the prediction of the pure
  bremssthalung for different power--law approximations for the Gaunt
  factor (short--dashed, dotted and long--dashed curves).}
\vspace{0.2truecm}
%\label{fi:mfps}
%\end{figure}

%\begin{multicols}{2}
%%%%%%%%%%%%%%

Therefore, given the above recipe to convert masses into luminosities,
each model is specified by four parameters, namely
$\sigma_8,\Omega_0,\alpha$ and $A$. The first two parameters specify
the cosmological scenario, while the others are linked to the
thermodynamics of the ICM. We follow a two--step procedure to place
constraints on such parameters:
\begin{description}
\item[(a)] for each value of $\Omega_0$ we constrain $\Gamma$, or
  equivalently $\sigma_8$, and $\alpha$ with the local XLF;
\item[(b)] we use XLF at $z>0.3$, number counts and redshift
  distribution from RDCS to constrain $\Omega_0$ and $A$, which
  specify the fluctuation growth and the ICM evolution, respectively.
\end{description}

The amplitude of the $L_{bol}$--$T$ relation is taken to be $L_6=2.9$
which represents the best--fitting value to the data by White et al.
(1997) and it is also consistent with the data by Arnaud \& Evrard
(1998) for clusters without cooling flows.  The remaining parameters
to be determined are $A$, which is connected to the thermodynamics of
the ICM, and $\Omega_0$ (along with $\Omega_\Lambda$), which
determines the growth rate of fluctuations.  The results will be
presented as constraints on the $\Omega_0$--$A$ parameter space.

%%FIGURE 3%%%
%\end{multicols}
%\begin{figure}
\includegraphics{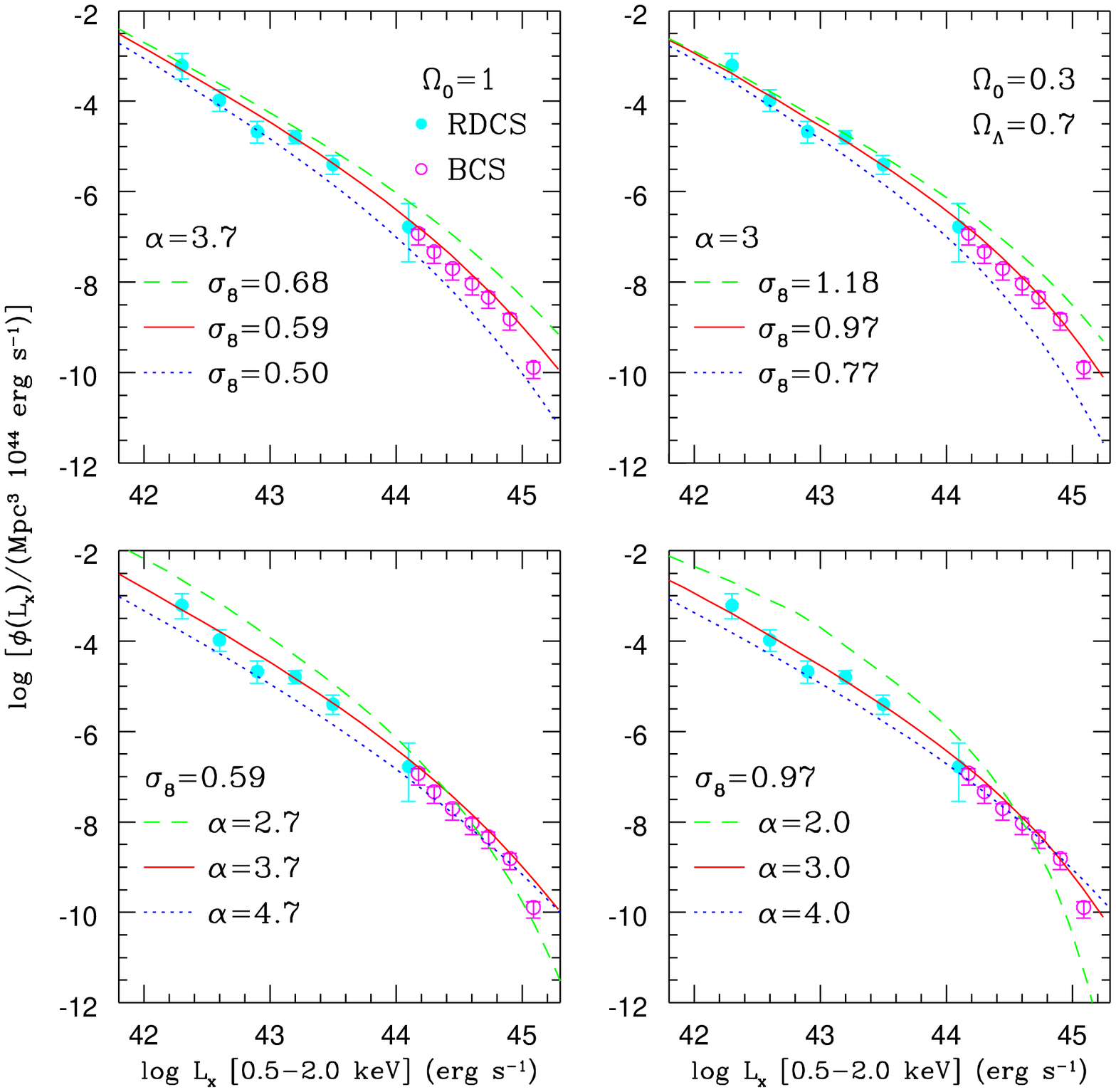}
%\special{psfile=f3.eps angle=0 voffset=-330 hoffset=-10 vscale=45 hscale=45}
$\ \ \ \ \ \ $\\
\vspace{8.2truecm}
$\ \ \ $\\
%\vspace{-0.5truecm}
{\small\parindent=3.5mm {Fig.}~3.--- 
The local XLF from RCDS (filled circles) and from
  BCS (open circles) are compared to model predictions. Left and right
  panels are for an $\Omega_0=1$ and a flat $\Omega_0=0.3$ model
  respectively. Upper panels show the effect of changing the shape
  parameter $\Gamma$ and, therefore, the $\sigma_8$ amplitude. Lower
  panels show the effect of changing the shape of the $L_{bol}$--$T$
  relation.}
\vspace{0.2truecm}
%\label{fi:mfps}
%\end{figure}

%\begin{multicols}{2}
%%%%%%%%%%%%%%

\section{Results from the local XLF}
The XLF in the soft {\sl ROSAT} energy band [0.5-2.0] keV 
is related to the PS mass distribution according to
\be
\phi(L)\,dL\,=\,{dn(M)\over dM}\,{dM\over dL}\,dL\,,
\label{eq:xlf}
\ee 
where for simplicity $L$ represents the luminosity in the
corresponding band.  In their paper, RDNG provided the XLF for the
RDCS sample at different redshifts for luminosities already converted
to the local rest--frame band by applying the appropriate
$K$--correction. We refer the reader to this paper for more details about 
the XLF computation.

In Figure 3 we show an example of how the XLF can be used
to place constraints on the model parameter space. We plot the local
XLF from the RDCS (filled circles) and the BCS at
$L_X>10^{44}\lum$ (Ebeling et al. 1997; open circles) in order to
cover the bright end of the XLF, not probed by the RDCS. We also show
different model predictions to emphasize the effect of changing the
spectrum amplitude $\sigma_8$ (upper panels) and the shape of the
$L_{bol}$--$T$ relation $\alpha$ (lower panels) for a critical density
model (left panels) and a flat low--density model with $\Omega_0=0.3$
(right panels). As expected, the effect of changing $\sigma_8$ in the
XLF is similar to that in the mass function: a rapid change in the
exponential cut--off with a smaller and smaller effect in the
low--luminosity tail. A steepening of the $L_{bol}$--$T$
relation (i.e., a larger $\alpha$)
corresponds to a larger luminosity range for a fixed temperature
range. As a result, the corresponding $\phi(L)$ becomes shallower.

Constraints on the amplitude of the power spectrum (or, equivalently,
the shape parameter $\Gamma$) and the slope of the local
$L_{bol}$--$T$ relation are placed by adopting a two--parameter
$\chi^2$--minimization procedure. The estimate of the $\chi^2$ between
model predictions and data is performed by assuming a log--normal
distribution of observational uncertainties in $\phi(L)$.  The
probability for a model to be accepted is then computed as the
probability that the data come from the parent model distribution with
Gaussian random variations in logarithmic units given by the size of
the error-bars. Note that
the BCS and the local RDCS XLFs have two slightly different
median redshifts, $\lb z\rb \simeq 0.1$ and $\lb z\rb \simeq
0.17$, respectively. This leads to a $\sim 6\%$ ambiguity when
constraining the amplitude of the power spectrum and, therefore,
$\sigma_8$. 

We show in Figure 4 the results from the local XLF. For
both the $\Omega_R=0$ and $\Omega_\Lambda =0$ cases (left and right
panels, respectively) we plot the $\Omega_0$ dependence for the shape
parameter $\Gamma$, the corresponding $\sigma_8$ from the {\sl COBE}
normalization and the slope $\alpha$ of the $L_{bol}$--$T$ relation.
The shaded regions represent the 90\% c.l. for model rejection. The
90\% c.l. for each one of the two fitting parameters has been
determined by fixing the other parameter at its best--fitting value
(i.e., minimum $\chi^2$). 
The dashed lines indicate the minimum--$\chi^2$ parameters if
$\beta =1$ is assumed instead of the fiducial value $\beta=1.15$.

As for $\sigma_8$ and $\Gamma$, we stress that these two quantities are
connected by a one-to-one relation, once the power--spectrum is {\sl
COBE}--normalized. As we show in Fig. 4, they are
determined with rather small uncertainties, thus confirming that the
cluster abundance provides a stringent constraint on the amplitude of
mass fluctuations at the cluster scale. 
The $\sigma_8$--$\Omega_0$ relation plotted in Fig.
4 can be analytically fitted as:
\ba 
\sigma_8 & = & (0.58\pm 0.02\pm 0.04)\times
\Omega_0^{-0.47+0.16\Omega_0} ~~; \nonumber \\
& & \Omega_\Lambda=1-\Omega_0 \nonumber \\
\sigma_8 & = & (0.58\pm 0.02\pm 0.04)\times
\Omega_0^{-0.53+0.27\Omega_0} ~~; \nonumber \\
& &\Omega_\Lambda=0
\label{eq:sigom}
\ea 
The first uncertainty corresponds to the 90\% c.l. from the
$\chi^2$ minimization, whereas the second error reflects the lack of a
common median redshift of the BCS and the local RDCS XLFs. 
Systematic uncertainties due to variations of $\delta_c$ with respect
to the canonical spherical--collapse value are not included in
eq.(\ref{eq:sigom}). However, both our analysis of $N$--body
simulations presented and the most recent
results by other authors (e.g., Governato et al. 1998; White S.D.M.,
private communication),
converge to indicate that such uncertainties are smaller than the
errors quoted in eq.(\ref{eq:sigom}).

%%FIGURE 4%%%
%\end{multicols}
%\begin{figure}
\includegraphics{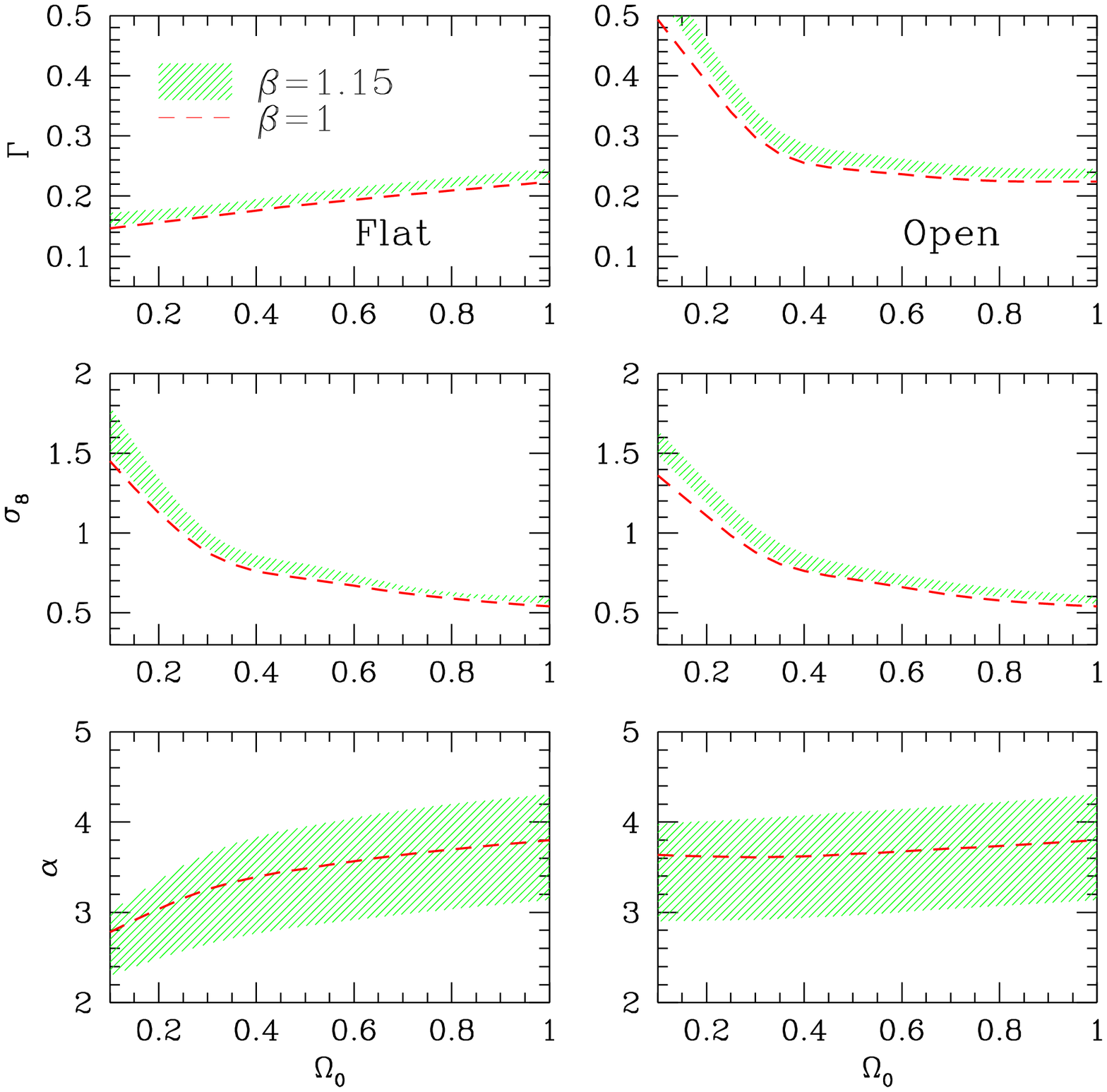}
%\special{psfile=f4.eps angle=0 voffset=-330 hoffset=-10 vscale=45 hscale=45}
$\ \ \ \ \ \ $\\
\vspace{8.2truecm}
$\ \ \ $\\
%\vspace{-0.5truecm}
{\small\parindent=3.5mm {Fig.}~4.--- 
Constraints on the model parameters from the local
  XLF, by combining the RDCS and the high--luminosity
  ($L_{[0.5-2]}>10^{44}\lum$) BCS data. Left and right panels refer to
  flat ($\Omega_R=0$) and open ($\Omega_\Lambda=0$) cases,
  respectively. The shaded areas indicate the 90\% c.l. as deduced
  from the $\chi^2$--minimization procedure, by assuming $\beta=1.15$
  (see text). The dashed curves indicates the best fitting parameters
  for $\beta=1$.}
\vspace{0.2truecm}
%\label{fi:mfps}
%\end{figure}

%\begin{multicols}{2}
%%%%%%%%%%%%%%

The above
scaling for $\sigma_8$ translates into an increasing trend for
$\Gamma$ if $\Omega_R=0$ and a decreasing trend for
$\Omega_\Lambda=0$. Such a behavior is due to the different
prescription for {\sl COBE} normalization of open and flat models.  For a
fixed shape of the power spectrum, flat models (e.g., Bunn \& White
1997) require values of $\sigma_8$ which are relatively larger and
larger than those required by open models (e.g., Hu \& White 1997) as
$\Omega_0$ decreases.  Therefore, in order to compensate for this
effects, open models select shallower spectra (i.e., larger $\Gamma$s)
at small $\Omega_0$. We also note that taking $\beta=1$ decreases the
central value in eq.(\ref{eq:sigom}) to $0.54$. This result agrees with
that found by Eke et al. (1996) who in fact assumed $\beta=1$, but
used the local X-ray cluster temperature function. 
Furthermore, any variation of $\delta_c$ with respect to its canonical value
turns into a proportional change of $\sigma_8$. 

Weaker constraints are instead obtained for the shape of the
$L_{bol}$--$T$ relation. For open model we find $3\mincir \alpha
\mincir 4$, roughly independent of $\Omega_0$, while flat models favor
somewhat smaller values at low $\Omega_0$. Such a difference is
required in order to compensate for the marginally larger $\sigma_8$,
for flat models at small $\Omega_0$, corresponding to a shallower XLF.
The acceptable $\alpha$ values cover the range of current
observational determinations of the $L_{bol}$--$T$ relation. 

\subsection{Effect of tilting the spectrum}
The results so far obtained from the local XLF assume a
scale--invariant primordial spectrum.  On the other hand, for a fixed
$\Omega_0$, the local XLF depends to a good approximation only on
$\sigma_8$. Therefore, for $n_{pr}\ne 1$, the shape parameter $\Gamma$
should be varied so as to leave $\sigma_8$ unchanged.  In this
respect, all the constraints on $A$ and $\Omega_0$ that we will
provide in the following are essentially independent of the assumption
of a Harrison--Zeldovich primordial spectrum. In Figure 5
we show how $\Gamma$ must be varied with $n_{pr}$,
for different $\Omega_0$, so as to keep $\sigma_8$ fixed at the
best--fitting value reported in Fig. 4. For flat models
we show the $\Omega_0=0.1,0.3,0.5,1$ cases, from lower to upper
curves, while $\Omega_0=0.2,0.3,0.5,1$ cases are shown for open
models, from upper to lower curves (a vanishing contribution from
tensor mode fluctuations to the CMB anisotropy is always assumed).
For instance, if $\Omega_0=1$, changing $n_{pr}=1$ to 0.8 implies that
the shape parameter must be increased from $\Gamma\simeq 0.24$ to
$\simeq 0.36$ in order to provide an equally good fit to the local
XLF. The shaded region indicates the 95\% c.l. constraint,
$\Gamma=0.23- 0.28(1-1/n_{pr})$ with 15\% uncertainty, obtained by
Liddle et al. (1996) from the shape of the APM galaxy power spectrum.
For flat models we find that $n_{pr}\magcir 1$ would require
$\Omega_0\magcir 0.5$, while a lower density parameter implies
$n_{pr}\mincir 1$. For open models, a blue $n_{pr}>1$ spectrum is
required by $0.3\mincir \Omega_0\mincir 1$ with a rapidly increasing
$n_{pr}$ for lower $\Omega_0$ values.
%%FIGURE 5%%%
%\end{multicols}
%\begin{figure}
\includegraphics{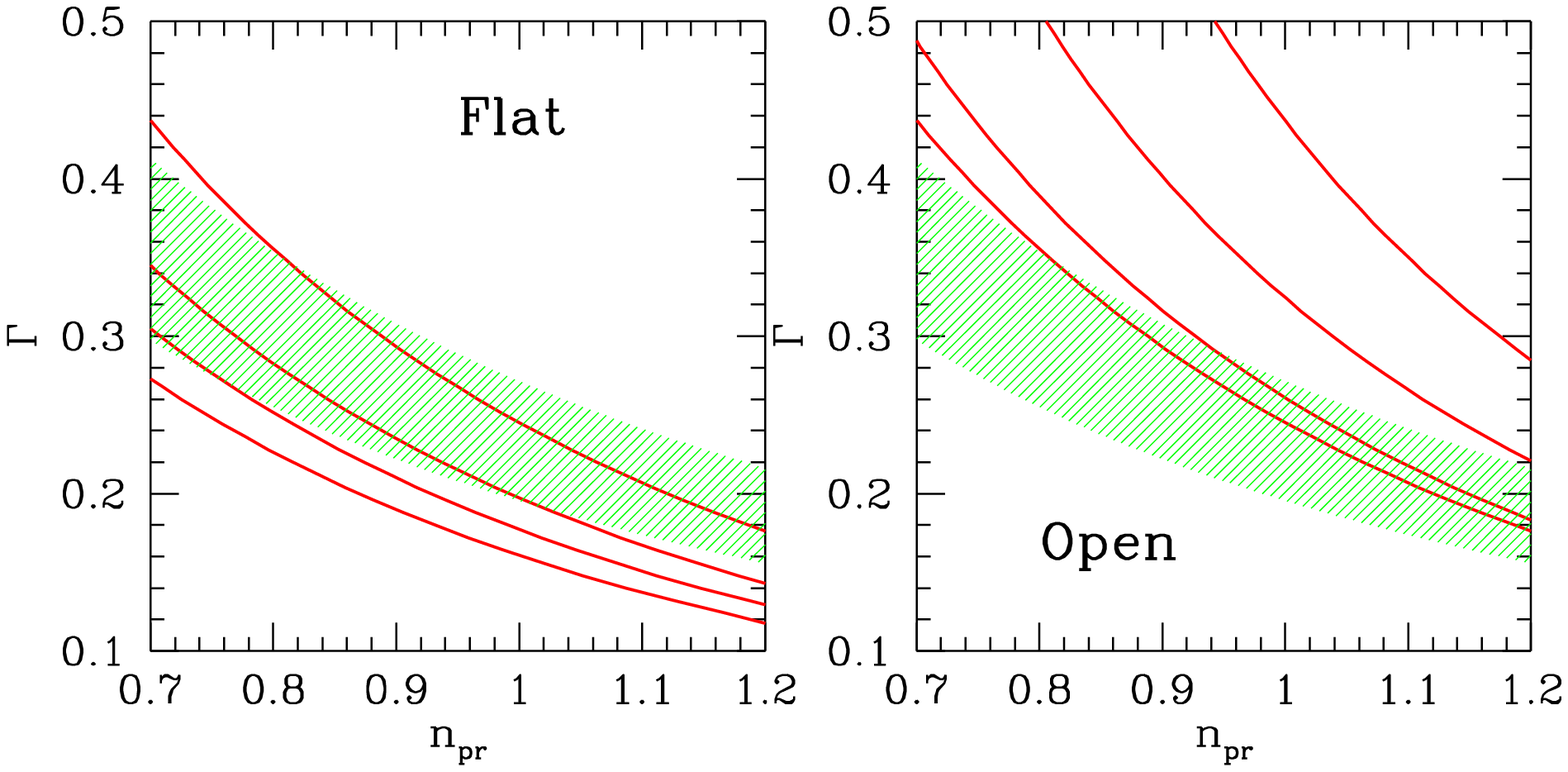}
%\special{psfile=f5.eps angle=0 voffset=-215 hoffset=-100 vscale=45 hscale=45}
$\ \ \ \ \ \ $\\
\vspace{4.4truecm}
$\ \ \ $\\
%\vspace{-0.5truecm}
{\small\parindent=3.5mm {Fig.}~5.--- 
The constraints from the local XLF on the shape
  parameter $\Gamma$ when the primordial spectral index $n_{pr}$ takes
  values different from unity. Each solid curve provides the
  best--fitting $\Gamma$s for different $\Omega_0$. For flat models
  (left panel) results are provided for $\Omega_0=0.1,0.3,0.5,1$ from
  lower to upper curves, while $\Omega_0=0.2,0.3,0.5,1$ are reported
  for open models (right panel), from upper to lower curves. The
  shaded area represents the 95\% c.l. constraint from the power
  spectrum of APM galaxies (Liddle et al. 1996).}
\vspace{0.2truecm}
%\label{fi:mfps}
%\end{figure}

\section{Tracing the redshift evolution}
In this section we will describe how the evolution of
the XLF, the number counts and the redshift distribution can be used
to place constraints in the $(\Omega_0,A)$ plane.

\subsection{The evolution of the XLF}
In Figure 6 we compare models predictions for two relevant cases with
the XLF data from EMSS by Henry et al. (1992) in the redshift bin
$z=[0.3-0.6]$ (median redshift $\lb z\rb=0.33$) and from the RDCS in
the redshift bins $z=[0.25-0.50]$ and $[0.50-0.85]$ (median redshifts
of $\lb z\rb=0.31$ and 0.60, respectively).  The XLF for different
redshift bins have been separated by $\Delta \log \phi(L)=2$ from each
other for reasons of clarity.  The EMSS XLF has been converted to the
0.5--2.0 keV band as described by RDNG.

%%FIGURE 6%%%
%\end{multicols}
%\begin{figure}
\includegraphics{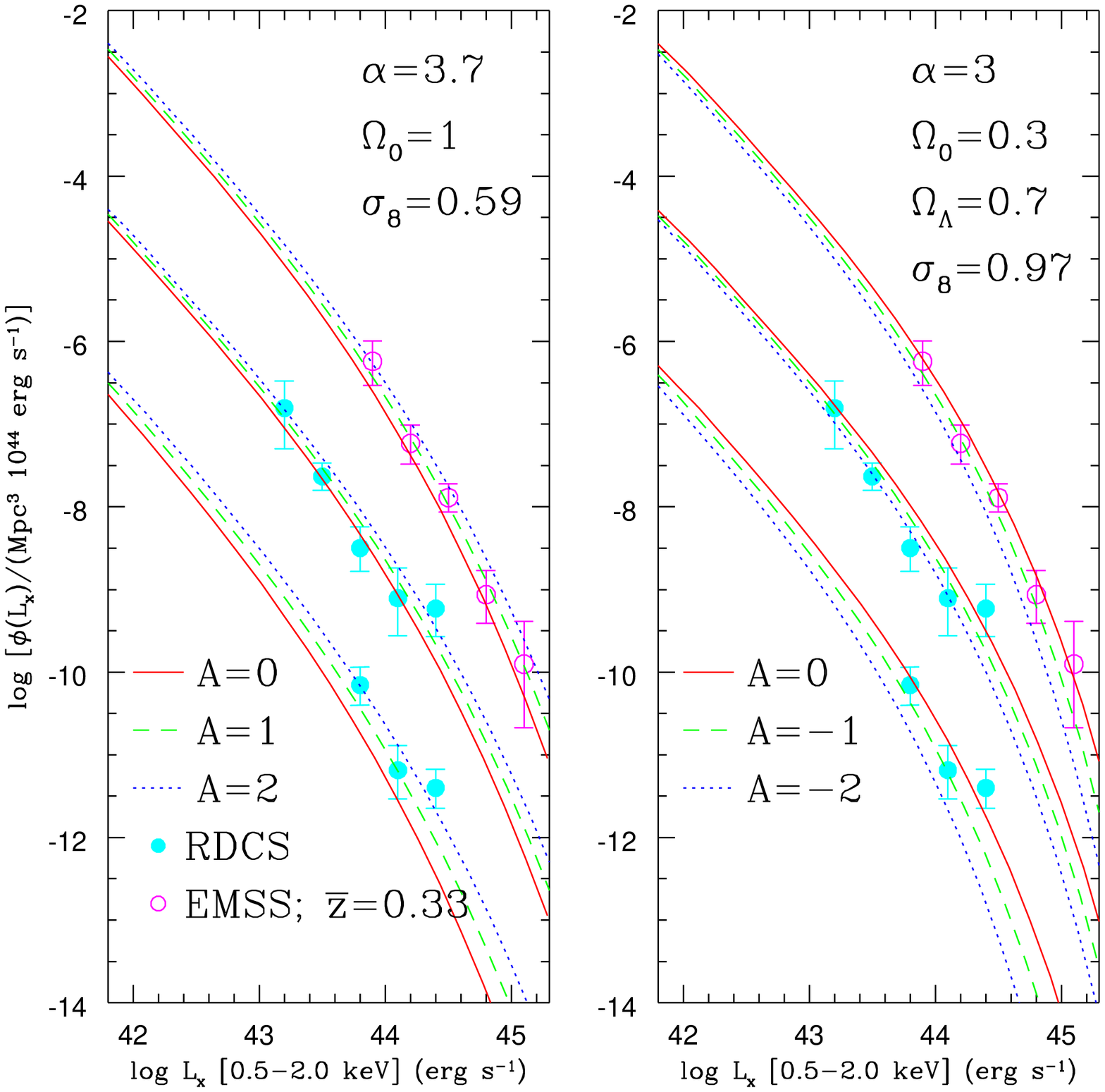}
%\special{psfile=f6.eps angle=0 voffset=-330 hoffset=-10 vscale=45 hscale=45}
$\ \ \ \ \ \ $\\
\vspace{8.2truecm}
$\ \ \ $\\
%\vspace{-0.5truecm}
{\small\parindent=3.5mm {Fig.}~6.--- 
The high--redshift XLF from EMSS (open circles)
  and from RCDS (filled circles) are
  compared to model predictions. The EMSS XLF refer to the redshift
  interval $z=[0.3-0.6]$, while that from RDCS refers to
  $z=[0.25-0.50]$ and $z=[0.50-0.85]$ respectively. The left and the
  right panels are for a $\Omega_0=1$ and a flat $\Omega_0=0.3$ model,
  respectively. For each model and at each redshift, different curves
  refer to different evolutions for the $L_{bol}$--$T$ relation.}
\vspace{0.2truecm}
%\label{fi:mfps}
%\end{figure}

In Figure 6 the shape of the
power spectrum is chosen so as to fit the local XLF (cf.
Fig. 4).  As expected, a positive evolution of the
$L_{bol}$--$T$ relation, i.e. an increase of the parameter $A$,
produces more luminous high--$z$ objects, thus an higher XLF.
The higher rate of evolution of the mass function in the $\Omega_0=1$
case requires a positive evolution of $L_{bol}$--$T$ in order to match
the lack of evolution of the XLF over a wide luminosity range.  For the
$\Omega_0=0.3$ case instead, the PS mass function evolves slowly at $z
\mincir 1$ (cf. Figure 1) so that a negligible or mild
negative $L_{bol}$--$T$ evolution is required by the RDCS XLF. A
somewhat more negative evolution seems to be required by the
steepening of the XLF observed from the EMSS data at $L_X \magcir
4\times 10^{44}\lum$, although the effect is statistically marginal.

We note that the absence of this steepening of the bright end of the
XLF at high redshifts would imply a more positive evolution of the
$L_{bol}$--$T$ relation. For instance, we verified that assuming a non
evolving XLF over the whole $L_X$ range would require
$A\simeq 2$ for an $\Omega_0=1$ model in Fig. 6, and
$A\simeq 0.5$ for the $\Omega_0=0.3$ flat model.

\subsection{The number counts}
We now consider the differential cluster number counts, $n(S)$, i.e.
 the number of objects per steradiant with flux in the range
$[S,S+dS]$. This observable can be computed starting from
eq.(\ref{eq:ps}) according to
\be
n(S)dS\,=\,\left({c\over H_0}\right)^3\int_0^\infty dz \,{r^2(z)\over
E(z)}\, n[M(S,z);z]\,{dM\over dS}\,dS
\label{eq:ns}
\ee
(cf. Kitayama \& Suto 1997)
where $r(z)$ is the radial coordinate appearing in the
Friedmann--Robertson--Walker metric:
\ba
r(z) & \!=\! & \int_0^z dz \,E^{-1}(z) ~~~~~;~~~~~\Omega_\Lambda=1-\Omega_0 \nonumber \\
r(z) & \!=\! & {2\left[\Omega_0z+(2-\Omega_0)\,(1-\sqrt{1+\Omega_0z})\right]\over
  \Omega_0^2(1+z) } ~;~\nonumber \\ & &\Omega_\Lambda=0\,.
\label{eq:rz}
\ea
The flux $S$ in the {\sl ROSAT} band is related to the luminosity
according to  
\be
S\,=\,{L\over 4\pi d_L^2(z)}\,,
\ee
where $d_L(z)=r(z)(1+z)$ is the luminosity distance at redshift $z$.
RDNG provided $n(S)$ for $S>2\times 10^{-14}\fl$ drawn from a
sample which is a factor two deeper than the one used to compute the
XLF. Here we consider a further extension of the Log N--Log S to the survey
flux limit. The large error bar of the faintest data point is due to 
incomplete optical identification at these fluxes.

%%FIGURE 7%%%
%\end{multicols}
%\begin{figure}
\includegraphics{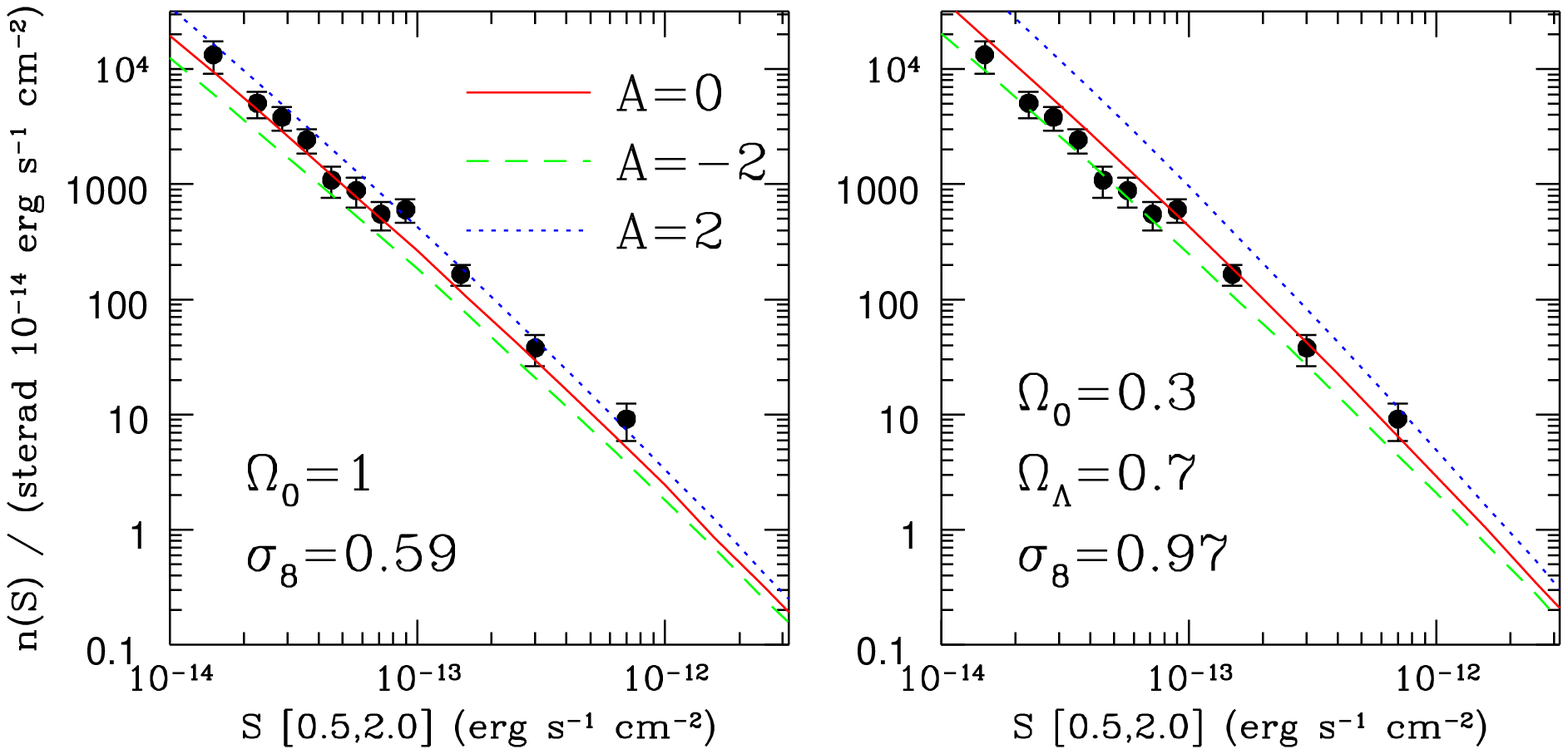}
%\special{psfile=f7.eps angle=0 voffset=-215 hoffset=-10 vscale=45 hscale=45}
$\ \ \ \ \ \ $\\
\vspace{4.2truecm}
$\ \ \ $\\
%\vspace{-0.5truecm}
{\small\parindent=3.5mm {Fig.}~7.--- 
The effect of the $L_{bol}$--$T$ evolution on the
  differential number counts, $n(S)$, for the same cosmological models
  shown in Figure 6.}
\vspace{0.2truecm}
%\label{fi:mfps}
%\end{figure}

In Figure 7 we show the effect of varying the evolution
of the $L_{bol}$--$T$ relation on $n(S)$ for the same models as in
Fig. 6. Again, a larger value of $A$ turns into an
increase of the flux coming from clusters at a given redshift, so as
to increase the number counts. Quite similar to the results from
the XLF, the critical--density model favors a weak positive
evolution while the low--density model requires a negative evolution
for $L_{bol}$--$T$.

\subsection{The redshift distribution}
The differential redshift distribution, $n(z)$, is defined as the
number of clusters above a given flux limit, in the effective survey area
of the RDCS, with redshift in the range $[z,z+dz]$. Its expression is
derived similarly to eq.(\ref{eq:ns}) and reads 
\ba
n(z)dz & \!=\! & \left({c\over H_0}\right)^3\,{r^2(z)\over
  E(z)}\,\int_{S_{lim}}^\infty dS\,f_{sky}(S)\nonumber \\
& \!\times \!&n[M(S,z);z]\,{dM\over
  dS}\,dz\,.
\label{eq:nz}
\ea 
In the above expression $S_{lim}=4\times 10^{-14}\fl$ is the limiting
flux above which the RDCS redshift distribution is complete, while
$f_{sky}(S)$ is the effective flux--dependent sky--coverage appropriate
for the RDCS sample (see Fig. 1 in RDNG). The upper panels of Figure
8 show the dependence of $n(z)$ on the evolution of the
$L_{bol}$--$T$ relation for the same two models previously considered.
In this case, the high--$z$ tail of the redshift distribution 
is highly sensitive to variations of the parameter $A$.  The rather extended
tail of the RDCS $dn(z)/dz$ rules out a negative $A$, if $\Omega_0=1$.
On the other hand, a flat model with $\Omega_0 \simeq 0.3$ requires
a mildly negative evolutionary index in order not to overproduce
high--redshift clusters.

%%FIGURE 8%%%
%\end{multicols}
%\begin{figure}
\includegraphics{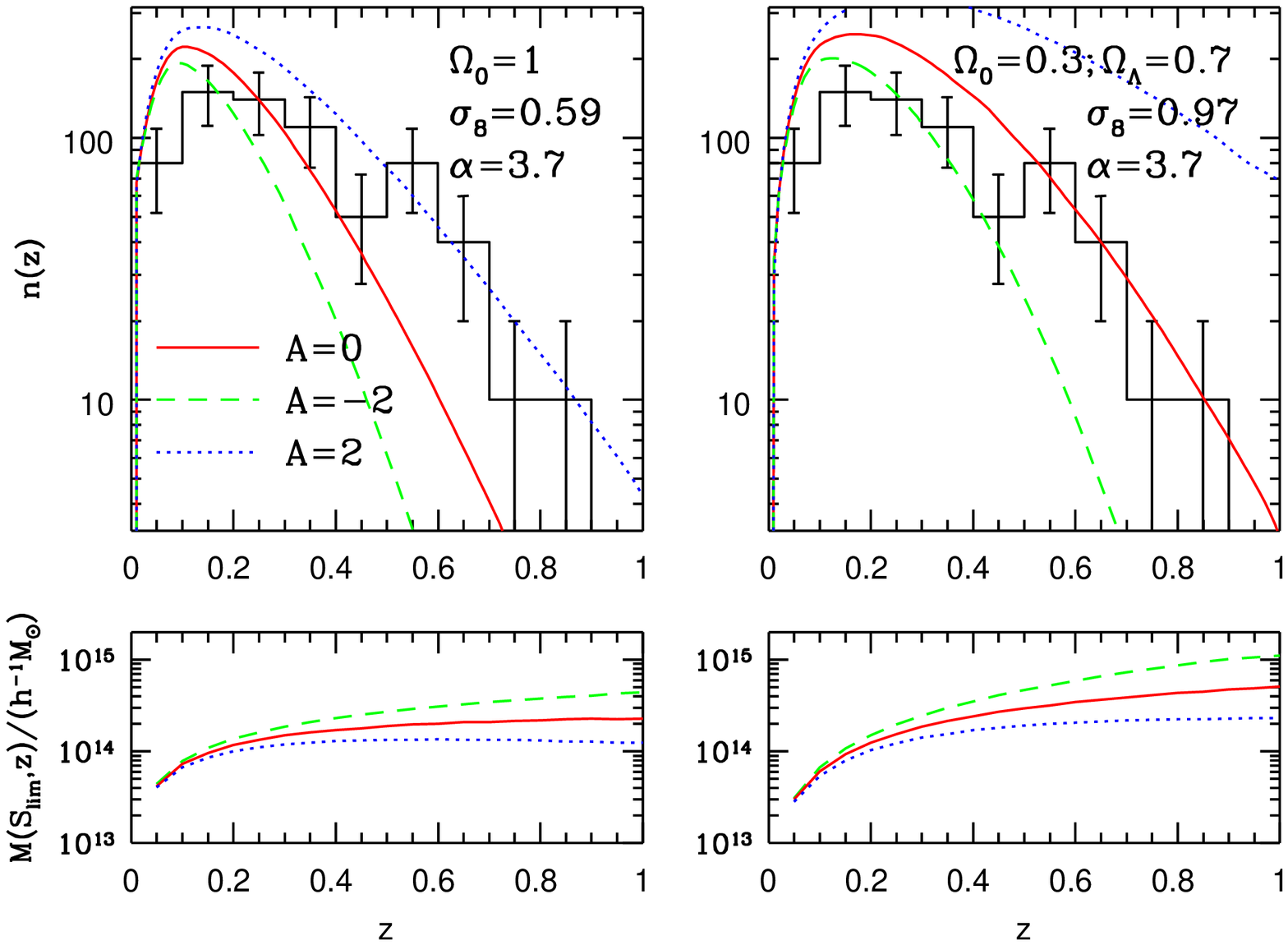}
%\special{psfile=f8.eps angle=0 voffset=-330 hoffset=-10 vscale=45 hscale=45}
$\ \ \ \ \ \ $\\
\vspace{6.2truecm}
$\ \ \ $\\
%\vspace{-0.5truecm}
{\small\parindent=3.5mm {Fig.}~8.--- 
The effect of the $L_{bol}$--$T$ evolution on the
  $n(z)$ redshift distribution, for the same cosmological models shown
  in Figure 6 (upper panels). The lower panels show for
  the same cases the mass of the smallest cluster that, at each
  redshift $z$, has a luminosity larger than $L_{lim}(z)=4\pi
  d_L^2(z)S_{lim}$ (where $S_{lim}=4\times 10^{-14}\fl$).}
\vspace{0.2truecm}
%\label{fi:mfps}
%\end{figure}

In the lower panels of Fig. 8 we also plot the mass corresponding to
the flux limit $S_{lim}$ as a function of the redshift for the
different choices of the $L_{bol}$--$T$ evolution.  This quantity
corresponds to the smallest mass involved in the Press--Schechter
computation of $n(z)$. Two aspects of this plot should be emphasized.
Firstly, assuming a negative $A$, larger masses at high $z$ are
included in order to keep the flux constant; {\em vice versa}, for
$A>0$ the luminosity at a fixed mass increases with $z$ and therefore,
much smaller masses at high redshift are required for a cluster to
emit above the RDCS flux limit. Secondly, such a minimum mass already
at $z\magcir 0.1$ keeps values at which the Press--Schechter formula
is succesfully tested against N--body simulations (cf. Fig.1). This
means that the small--mass regime where the PS approach has been shown
to overpredict the cluster abundance should not affect the results of
our analysis.

\subsection{Constraints on the $A$--$\Omega_0$ plane}
In order to estimate the range of $A$ values allowed for each model
we apply a one--parameter $\chi^2$ minimization. For each 
$\Omega_0$, the values of $\sigma_8$ and $\alpha$ are
those corresponding to the best--fit to the local XLF (cf. the
previous section).  A global summary of the resulting constraints on
the $\Omega_0$--$A$ plane from the three observational constraints is
shown in Figure 9. For both flat (left panels) and open
(right panels) models, the shaded areas correspond to the 90\% c.l.
for $A$. As a general result, these findings confirm on a more
quantitative ground that $\Omega_0\simeq 1$ models prefer a positive
$L_{bol}$--$T$ evolution, while lower $A$ values are
required by smaller $\Omega_0$. 

%%FIGURE 9%%%
%\end{multicols}
%\begin{figure}
\includegraphics{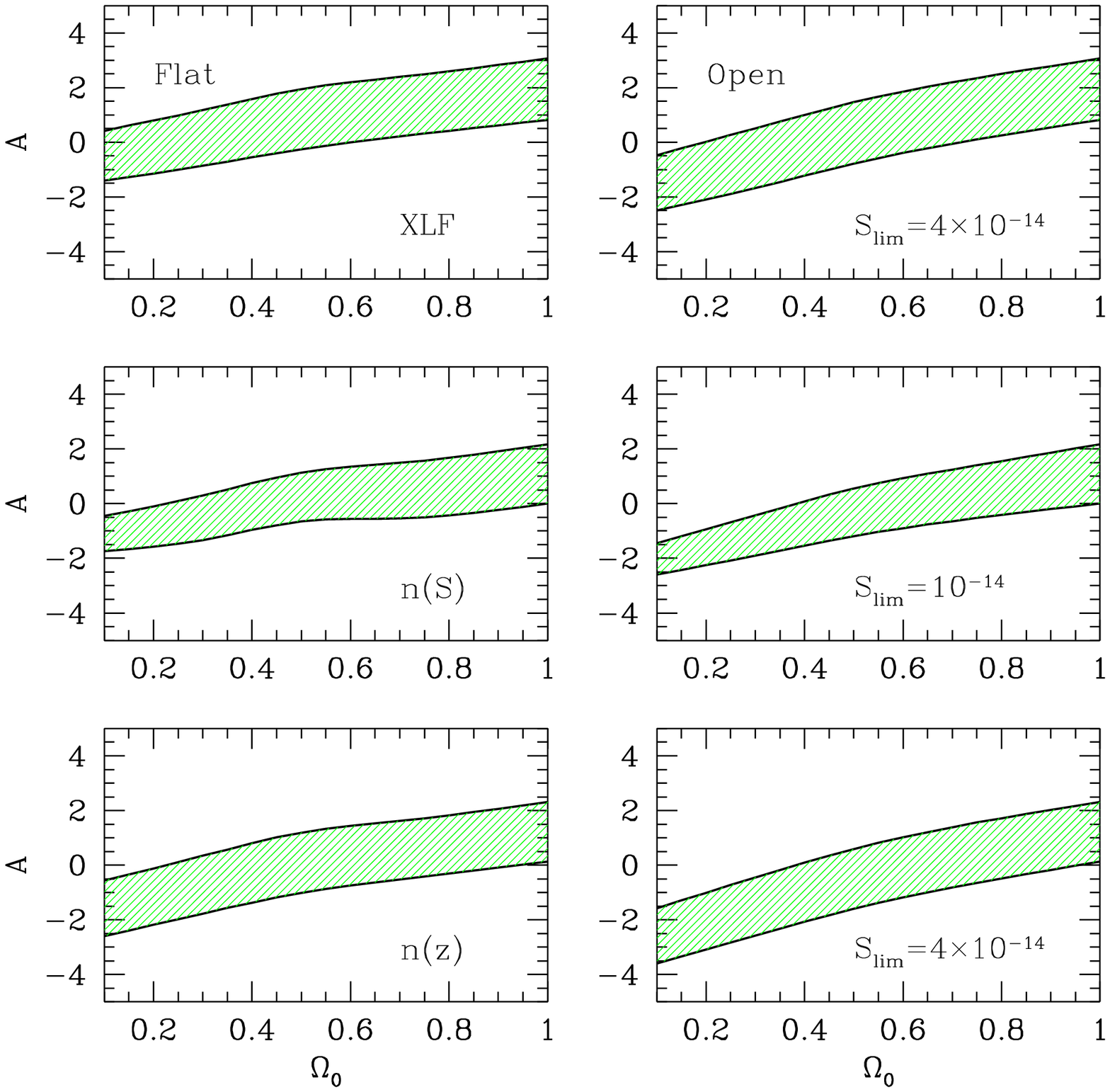}
%\special{psfile=f9.eps angle=0 voffset=-330 hoffset=-10 vscale=45 hscale=45}
$\ \ \ \ \ \ $\\
\vspace{8.2truecm}
$\ \ \ $\\
%\vspace{-0.5truecm}
{\small\parindent=3.5mm {Fig.}~9.--- 
Constraints at the 90\% c.l. on the $\Omega_0$--$A$
  plane from the $z>0.3$ XLF, number counts $n(S)$ and redshift
  distribution $n(z)$ (from upper to lower panels, for both flat and
  open models (left and right panels respectively).  The limiting
  fluxes of the samples (in units of $\fl$) from which the XLF, $n(S)$
  and $n(z)$ are derived are also indicated.}
\vspace{0.2truecm}
%\label{fi:mfps}
%\end{figure}

As expected, the constraints provided by XLF, $n(S)$ and $n(z)$ are
almost coincident, although the XLF tends to prefer marginally larger
values of $A$. Small differences between
such constraints may be expected in principle.  Indeed, they come from
somewhat different samples. The XLF constraint includes the EMSS
results for $L_{[0.5,2]}$ at $\lb z\rb=0.33$, as well as the RDCS data
above $S_{lim}=4\times 10^{-14}\fl$. The $n(S)$, instead, has been
computed for RDCS clusters down to $S_{lim}=1\times 10^{-14}\fl$, while
$n(z)$ is based on the same flux--limit as the XLF. 

A tentative analytic fit of the constraints on the $\Omega_0$--$A$
plane from the redshift distribution is provided by the expressions 
\ba 
A\,=\,(3\Omega_0-2)\pm 1
~~~~~~;~~~~~~\Omega_R=0\,;\nonumber \\ A\,=\,(4\Omega_0-3)\pm 1
~~~~~~;~~~~~~\Omega_\Lambda=0\,.  
\ea 
A similar result has recently been obtained, only for the open models,
by Sadat et al. (1998) using the redshift distribution of the EMSS
sample. As a general result, it is evident how
an independent determination of the evolution
of the $L_{bol}$--$T$ relation and a deeper understanding of
the nature of its scatter can turn into a constraint on the density
parameter. For instance, if
a redshift--independent $L_{bol}$--$T$ relation 
is assumed (i.e. $A=0$), then the evolution of the XLF implies
$\Omega_0=0.4^{+0.3}_{-0.2}$ if $\Omega_\Lambda=0$ and $\Omega_0 \le
0.6$ if $\Omega_R=0$. These findings are consistent with those
obtained by Eke et al. (1998) from the analysis of the $X$--ray
temperature function at $z\mincir 0.4$,
also based on a non--evolving $L_{bol}$--$T$ relation.

As for the physical meaning of the evolution parameter $A$, detailed
models involving both the gravitational shock heating and
non--gravitational heating (see Cavaliere, Menci \& Tozzi 1997, Tozzi
\& Norman in prep.), predict a range of $A$ values between 0 and 1.5
at the cluster mass scale, and slightly lower values, $-0.5\mincir A
\mincir 1$, for groups.  This result holds if the non--gravitational
heating rate has a smooth dependence on the redshift, as suggested by
the average star formation rate derived from deep galaxy surveys
(e.g., Madau et al. 1996). However, the duration and the epoch of the
heating of the ICM, which is not well known at present, critically
affects the effective value of $A$, so that a self consistent modeling
of the feedback processes from galaxies is needed in order to
constrain A within a narrow range.  This justifies our present choice
to leave $A$ as a parameter free to be fit by the evolution of the
XLF.

\section{Conclusions}
In this paper we have used observational results from the Rosat Deep
Cluster Survey by Rosati et al. (1998) to place constraints
on cosmological models. The unprecedented extension of the RDCS both
in redshift ($z\mincir 0.8$) and fluxes ($S\magcir 1\times
10^{-14}\fl$ in the [0.5-2.0] keV band) makes it, in principle, a 
suitable baseline over which the evolutionary pattern of the cluster
abundance and, therefore, the density parameter $\Omega_0$, can be
investigated. A necessary ingredient for this analysis is the physics
of the intra--cluster gas, which drives the relation between the
distribution of cluster halo masses and the observed distribution of
$X$--ray luminosities. To this purpose, we adopt a parametrical
approach in which the shape of the local $L_{bol}$--$T$ relation,
its evolution and the parameters specifying the cosmological models
are {\em all} fitted against observational data.

Firstly, we have used the local XLF from RDCS and the extension at high
luminosities from the XLF of the Brightest Cluster Sample (Ebeling et
al. 1997) to fix the r.m.s. fluctuation amplitude at the cluster mass
scale and the shape of the local $L_{bol}$--$T$ relation. Secondly,
we have used the evolving XLF, the flux number counts and the redshift
distribution with the aim of constraining the density parameter
$\Omega_0$ and the evolution of the $L_{bol}$--$T$ relation.

The main results of our analysis can be summarized as follows.

\begin{description}
\item[(a)] A careful comparison between the PS predictions and N--body
  simulations confirms that this analytical approach for the
  distribution of virial halo masses is adequate over a rather large
  mass range around the non--linear mass scale. The low mass tail,
  where the PS formula overpredicts the halo abundance, has been shown
  to lie outside the mass range probed by RDCS clusters. 

\item[(b)] The local XLF data constrain the amplitude of the power
spectrum according to 
\ba 
\sigma_8 &\!= \!& (0.58\pm 0.06)\times
\Omega_0^{-0.47+0.16\Omega_0} ~~~~; \nonumber \\
& & \Omega_\Lambda=1-\Omega_0
\nonumber \\ 
\sigma_8 &\!= \!& (0.58\pm 0.06)\times
\Omega_0^{-0.53+0.27\Omega_0} ~~~~; \nonumber \\ 
& & \Omega_\Lambda=0\,.
%\label{eq:sigom} 
\ea 
This result agrees with analyses of the $X$--ray
temperature function (e.g., Viana \& Liddle 1996; Eke et al. 1996;
Markevitch 1998), the optical virial mass function (e.g., Girardi et
al. 1998) and the $X$--ray cluster number counts alone.
  
  As for the shape of the local $L_{bol}$--$T$ relation, we find
  $3\mincir \alpha \mincir 4$, quite independent of the cosmological
  model and in general agreement with observational results (e.g.,
  David et al. 1993; White et al. 1997).
  
\item[(c)] From the observed evolution of the XLF we constrain the
density parameter $\Omega_0$ and the evolution parameter $A$ of the
$L_{bol}$--$T$ relation. We find that $\Omega_0=1$ models require a
positive evolution of the $L_{bol}$--$T$ relation with $1\mincir
A\mincir 3$.  A non--evolving $L_{bol}$--$T$ ($A=0$), consistent with
data at $z\mincir 0.4$ (Mushotzky \& Scharf 1997), implies
$\Omega_0=0.4^{+0.3}_{-0.2}$ for open models and $\Omega_0\mincir 0.6$
for flat models.

\end{description}

This paper set out to address the two principal issues given in the 
Introduction. In summary, the above results lead us to conclude that:
(1) Available data on the local cluster XLF place 
robust constraints on $\sigma_8$ as a function of $\Omega_0$ [cf. 
eqs.(\ref{eq:sigom})], without any {\em a priori} assumption about the 
local $L_{bol}$--$T$ relation and; (2) 
Present uncertainties in the evolution of the 
$L_{bol}$--$T$ relation do not as yet give strong constraints on 
$\Omega_0$, even with a high--redshift sample as deep as RDCS.

Samples like RDCS
will represent a fundamental basis for the study of distant clusters
in the years to come. The next generation of $X$--ray satellites and
the already available large optical telescopes should open the
possibility of determining cluster masses via $X$--ray temperature
measurements, virial analysis and gravitational lensing
studies. Carrying on such observations for an even limited number of
clusters, extracted from a well defined statistical sample, will
determine both the evolution of the cluster internal dynamics and the
value of the cosmological density parameter.

\acknowledgments S.B. wishes to acknowledge the European Southern
Observatory and the Johns Hopkins University for hospitality during
several phases of preparation of this work. We wish to thank Hugh
Couchman for the generous sharing of his adaptive P3M code. We are
also grateful to Alfonso Cavaliere and Riccardo Giacconi for
stimulating discussions and continuous encouragement on this work, and
to Pat Henry for a careful reading of the paper. P.R. thanks the
Telescope Allocation Committees of Kitt Peak National Observatory,
Cerro Tololo Inter-American Observatory, and ESO for the allocation of
generous observing time. This work has been supported by NASA
grants NAG 8-1133 and NAG 5-3537.

%%%%%%%%%%%% per formato preprint
\end{multicols}
%%%%%%%%%%%% per formato preprint

\end{document}